\documentclass[a4paper,notitlepage,12pt]{article}

\begin{document}
\title{Thermocapillary migration of a droplet with a thermal source at large Reynolds and Marangoni numbers}
\author{ Zuo-Bing Wu\footnotemark[1]\\
State Key Laboratory of Nonlinear Mechanics,\\
Institute of Mechanics,\\ Chinese Academy of Sciences, Beijing
100190, China}
 \maketitle

\footnotetext[1]{Author to whom correspondence should be
addressed. Tel: 86-10-82543955; fax: 86-10-82543977. Email:
wuzb@lnm.imech.ac.cn(Z.-B. Wu).}

\newpage
\begin{abstract}
The {\it unsteady} process for thermocapillary droplet migration at large Reynolds and Marangoni
numbers has been previously reported by identifying a nonconservative integral thermal
 flux across the surface in the {\it steady} thermocapillary droplet migration, [Wu and Hu, J. Math. Phys. {\bf 54}
023102, (2013)]. Here we add a thermal source in the droplet to preserve the integral thermal
 flux across the surface as conservative, so that thermocapillary droplet migration
 at large Reynolds and Marangoni numbers can reach a {\it quasi-steady} process. Under assumptions
 of {\it quasi-steady} state and non-deformation of the droplet,
we make an analytical result for the {\it steady} thermocapillary migration of
droplet with the thermal source at large Reynolds and Marangoni numbers. The
result shows that the thermocapillary droplet migration speed
slowly increases with the increase of Marangoni number.


\end{abstract}

\newpage
\section{Introduction}
Thermocapillary migration of a droplet or bubble at a uniform temperature
gradient is a very interesting topic in both microgravity and microfluidics\cite{1,DT}.
On the one hand, due to the development of space exploration, the studies of
the physical mechanism of the droplet/bubble migration phenomena in the microgravity
environment become more and more important. The first investigation of thermocapillary
droplet migration was completed by Young et al (YGB) in 1959, who gave an analytical prediction for the
migration speed at zero limit Reynolds(Re) and Marangoni(Ma) numbers\cite{2}.
Subramanian\cite{3} proposed the quasi-steady state assumption for
thermocapillary bubble migration and
obtained analytical results with high order expansion  at small Ma numbers.
 Balasubramanian and Chai\cite{10} derived an exact
result for thermocapillary droplet migration at small Ma numbers. The experimental results for the
droplet migration speed at small Re numbers obtained by Braun et
al \cite{11} are in agreement with those of the YGB model. On the other hand, the manipulation and
actuation of droplets in microfluidic devices have an extensive application
on the chemical industry and biological engineering. Many interest problems,
such as static contact and dynamic control of the droplets on solid substrates,
were theoretically analyzed and experimentally observed\cite{Smith,CTDW,SKLP,NC}.

Although the thermocapillary droplet migration processes at small Ma numbers
under the microgravity environment are understood
very well in the series of theoretical analyses and experimental investigations,
the physical behaviors at large Ma numbers are rather complicated due to
the momentum and energy transfer though the interface of two-phase fluids.
In large Re and Ma number
area, it was reported\cite{12} that the migration speed of a droplet
increases with the increase of Ma number, as is in qualitative
agreement with the corresponding numerical simulation\cite{13}.
The above theoretical analysis and numerical simulation are based on
assumptions of the quasi-steady state and non-deformation of the
droplet.
The experimental investigation carried out by
Hadland et al\cite{9} and Xie et al\cite{14}
showed that the droplet migration speed
non-denationalized by the YGB velocity decreases with the increase
of Ma number. The results are not in
qualitative agreement with the above theoretical and numerical
ones.
Moreover, a nonconservative integral thermal
flux across the surface in the steady thermocapillary droplet migration
at large Ma(Re) numbers was identified\cite{WH3}. It was indicated that the thermocapillary
droplet migration at large Ma(Re) numbers is an unsteady process.
Therefore, the thermocapillary droplet migration at large Ma(Re) numbers remains
a topic to be studied with respect to its physical mechanism.

In this paper, first, under the assumption of the quasi-steady process of
thermocapillary droplet migration at large Re and Ma numbers,
a thermal source is added in the droplet to change the integral thermal
flux across the surface from nonconservative to conservative. Then, we make
an analytical result for the steady thermocapillary migration of droplet with the
thermal source at large Re and Ma numbers and
elucidate effects of the adding thermal source in the droplet on 
thermocapillary migration process.

\section{Problem formulation}
Consider
the thermocapillary migration of a spherical droplet of radius $R_0$,
density $\gamma \rho$, dynamic viscosity $\alpha \mu$, thermal
conductivity $\beta k$, and thermal diffusivity $\lambda \kappa$
in a continuous phase fluid of infinite extent with density
$\rho$, dynamic viscosity $\mu$, thermal conductivity $k$, and
thermal diffusivity $\kappa$ under a uniform temperature gradient
$G$. The rate of change of the interfacial tension between the
droplet and the continuous phase fluid with temperature is denoted by
$\sigma_T$. Figure 1 displays a schematic diagram of thermocapillary
migration of a droplet with a thermal source in a laboratory coordinate system
denoted by a bar.
Axisymmetric energy equations for the continuous phase
and the fluid within the droplet are written as follows
\begin{equation}
\begin{array}{l}
\frac{\partial{\bar{T}}}{\partial t} + \bar{\bf v} \bar{\nabla} \bar{T}= \kappa \bar{\Delta} \bar{T},\\
\frac{\partial{\bar{T'}}}{\partial t} + \bar{\bf v'} \bar{\nabla} \bar{T'}= \lambda \kappa \bar{\Delta} \bar{T'} + \Omega,
\end{array}
\end{equation}
where $\bar{\bf v}$ and $\bar{T}$ are velocity and temperature,
a prime and $\Omega$ denote quantities and a thermal source in the droplet.
By going over an
unstable migration process from beginning, the droplet migration can
reach a steady state, i.e., with the constant  migration speed
$V_{\infty}$. Using the coordinate transformation from the
laboratory coordinate system to a coordinate system moving with
the droplet velocity $V_{\infty}$
\begin{equation}
\begin{array}{lll}
\bar{\bf r} = {\bf r} + V_{\infty}t {\bf k},& \bar{\bf v}(\bar{\bf r},t) = {\bf v}({\bf r}) + V_{\infty} {\bf k},& \bar{T}(\bar{\bf r},t )= T({\bf r}) + GV_{\infty}t,\\
&  \bar{\bf v'} (\bar{\bf r},t)  = {\bf v'}({\bf r}) + V_{\infty} {\bf k},& \bar{T'}(\bar{\bf r},t )= T'({\bf r})  + GV_{\infty}t,
\end{array}
\end{equation}
the problem (1) can be formulated as
\begin{equation}
\begin{array}{l}
G V_{\infty} + {\bf v} \nabla T= \kappa \Delta T,\\
G V_{\infty} + {\bf v'} \nabla  T'= \lambda \kappa \Delta T' +\Omega.
\end{array}
\end{equation}
 By taking the radius of the droplet $R_0$, the YGB model velocity
$v_o=-\sigma_T G R_0/\mu$ and $GR_0$ as reference quantities to
make coordinates, velocity and temperature dimensionless, the
energy equations (3) can be written in the following dimensionless
form in the spherical coordinate system ($r, \theta$) as follows
\begin{eqnarray}
1+u \frac{\partial{T}}{\partial{r}} +\frac{v}{r} \frac{\partial{T}}{\partial{\theta}}= \epsilon^2 \Delta T,\\
1+u' \frac{\partial{T'}}{\partial{r}} +\frac{v'}{r} \frac{\partial{T'}}{\partial{\theta}}=\lambda \epsilon^2 \Delta T' + \frac{\Omega}{G v_0 V_\infty},
\label{1}
\end{eqnarray}
where the small parameter $\epsilon$ and Ma number are
defined, respectively, as
\begin{equation}
\epsilon=\frac{1}{\sqrt{Ma V_{\infty}}}
\end{equation}
and
\begin{equation}
Ma=\frac{v_0R_0}{\kappa}.
\end{equation}

In the space experiments\cite{9,14}, Prandtl numbers ($Pr=\frac{\mu}{\rho \kappa}$) are constants
in terms of the fixed fluid media, but Ma numbers are changed duo to the variable droplet sizes.
It is common to adopt Re numbers ($Re = \frac{\rho v_0 R_0}{\mu} = \frac{Ma}{Pr}$) instead of Pr numbers
in order to capture velocity fields with the changes of Re numbers. In general, the
deformation of droplet usually depends on the Weber numbers [$We=\frac{\rho v^2_0 R}{\sigma_0}=CaRe$, $Ca(=\frac{v_0 \mu}{\sigma_0})$
is the Capillary number], it is not included in this study
due to the small We numbers in the space experiments. For example,
$We$ is in $O(10^{-3}) - O(10^{-1})$ when $Ma < 1000$\cite{14,WH2}.
Thus, the solutions of Eqs. (4) and (5) have to satisfy the following
boundary conditions at the interface of two phase fluids ($r=1$)
\begin{eqnarray}
T(1,\theta) =T'(1,\theta),\\
\frac{\partial{T}}{\partial r}(1,\theta) = \beta \frac{\partial{T'}}{\partial r}(1,\theta)
\end{eqnarray}
and at places far away from the droplet
\begin{equation}
T \to r \cos \theta, {\rm as} \  r \to \infty.
\end{equation}
Besides the above temperature boundary conditions, the stress boundary conditions
at the interface are also applied for the steady droplet migration.
However, the normal stress boundary condition is removed in terms of
the non-deformation assumption. The shear stress
boundary condition is expressed by
\begin{eqnarray}
\tau_{r\theta} -\alpha \tau'_{r\theta} = \frac{1}{V_\infty r}  \frac{\partial T}{\partial \theta},
\end{eqnarray}
where
\begin{eqnarray}
\tau_{r\theta}= r \frac{\partial }{\partial r}(\frac{v}{r}) + \frac{1}{r} \frac{\partial u}{\partial \theta}.
\end{eqnarray}



For large Re numbers, the inner and outer momentum boundary layers
 at $r$-direction are introduced near the surface of droplet.
In the flow fields outside the momentum boundary layers,
there are still the potential flows. The scaled inviscid velocity field in the continuous
phase and Hill's spherical vortex within the droplet can be written as, respectively\cite{15}
\begin{equation}
\begin{array}{l}
U=-\cos \theta (1-\frac{1}{r^3}),\\
V=\sin \theta (1+\frac{1}{2r^3})
\end{array}
\end{equation}
and

\begin{equation}
\begin{array}{l}
U'=\frac{3}{2} \cos \theta (1-r^2),\\
V'=- \frac{3}{2} \sin \theta (1-2r^2).
\end{array}
\end{equation}
By introducing the perturbation velocity fields ${\bf \hat{v}} = {\bf v} - {\bf U}$, ${\bf \hat{v}'} = {\bf v'} - {\bf U'}$
and the depth of
boundary layer $\delta x =r-1$ $(\delta =\sqrt{\frac{1}{Re}})$
into the boundary condition (11) and the momentum equation for the continuous fluid, they are derived and truncated in
the leading orders $O(1/\delta)$ and $O(\delta)$ as, respectively\cite{15}
\begin{equation}
\frac{\partial \hat{v}}{\partial x} = \frac{3}{2}(2+3\alpha) \sin \theta + \frac{1}{V_\infty} \frac{\partial T}{\partial \theta}.
\end{equation}
and
\begin{equation}
\frac{3}{2} \frac{\partial}{\partial \theta} (\hat{v} \sin \theta) -3x \cos \theta \frac{\partial \hat{v}}{\partial x} = \frac{\partial^2 \hat{v}}{\partial x^2}.
\end{equation}
where $\frac{\partial}{\partial r}=O(\frac{1}{\delta})$, $U/U'=O(\delta)/O(\delta)$, $V/V'=O(1)/O(1)$,
$\hat{u}/\hat{u'}=O(\delta^2)/O(\delta^2)$ and $\hat{v}/\hat{v'}=O(\delta)/O(\delta)$ are in the momentum boundary layers.
Following the derivation for the steady migration speed of a bubble at large Re numbers\cite{CJ}, the migration speed of the droplet
is written as follows
\begin{equation}
V_\infty = -\frac{1}{2(2+3\alpha)} \int_0^{\pi} \sin^2 \theta \frac{\partial T}{\partial \theta} (1,\theta) d\theta
 = \frac{1}{2+3\alpha} \int_0^{\pi} \sin \theta \cos \theta T (1,\theta) d\theta.
\end{equation}
Thus, to determine $V_\infty$, the temperature $T(1,\theta)$ on the surface of droplet is
required. However, before the energy equations are solved, the self-consistency of
 temperature fields for the steady droplet migration system will be determined. It means that
the solutions of steady energy equations will satisfy the conservative
integral thermal flux across the surface of droplet.
In this case, the temperature field at infinity in Eq. (10) was derived by using
the asymptotic expansion method and expressed as\cite{WH3}
\begin{equation}
T \approx r \cos \theta -\frac{1}{2r^2} \cos \theta +o(1).
\end{equation}

Integrating Eqs. (4) and (5) in the continuous phase domain
$(r\in [1,r_{\infty}],\theta\in[0,\pi])$
 with boundary condition (18) and within
the droplet region $(r\in [0,1],\theta\in[0,\pi])$, respectively, we obtain

\begin{equation}
\int_0^{\pi} \frac{\partial{T}}{\partial{r}} (1,\theta) \sin \theta d \theta = -\frac{1}{3 \epsilon^2}\\
\end{equation}
and

\begin{equation}
\int_0^{\pi} \frac{\partial{T'}}{\partial{r}} (1,\theta) \sin
\theta d \theta = \frac{2}{3 \lambda \epsilon^2}
(1-\frac{\Omega}{G v_0 V_\infty}).
\end{equation}
From Eq. (19) and Eq. (20), we have
\begin{equation}
\begin{array}{l}
\beta \int_0^{\pi} \frac{\partial{T'}}{\partial{r}}
(1,\theta) \sin \theta d \theta - \int_0^{\pi}
\frac{\partial{T}}{\partial{r}} (1,\theta) \sin \theta d
\theta\\
= \int_0^{\pi} [ \beta \frac{\partial{T'}}{\partial{r}}
(1,\theta) -\frac{\partial{T}}{\partial{r}} (1,\theta) ] \sin \theta d \theta \\
= \frac{1}{3 \epsilon^2}
(1+ \frac{2 \beta}{\lambda})-\frac{2 \beta \Omega}{3 \lambda G v_0} Ma.
\end{array}
\end{equation}
 For large Ma numbers and finite $V_\infty$,
 Eqs. (19) and (20) should satisfy the thermal flux boundary condition (9),
 i.e., the right side of Eq. (21) will be zero. So, we have
\begin{equation}
\Omega = \frac{\lambda Gv_0 V_\infty}{2 \beta} (1 + \frac{2 \beta}{\lambda})
= Gv_0 V_\infty (1 + \frac{\lambda}{2 \beta}).
\end{equation}
Using Eq. (22), Eq. (5) is rewritten as
\begin{equation}
u' \frac{\partial{T'}}{\partial{r}} +\frac{v'}{r} \frac{\partial{T'}}{\partial{\theta}}=\lambda \epsilon^2 \Delta T' + \frac{\lambda}{2 \beta}.
\label{1}
\end{equation}
In following, we will focus on the
steady thermocapillary migration of droplet with the thermal source under a uniform
temperature gradient and determine the dependence of the migration
speed on large Ma numbers.

\section{Analysis and results}
\subsection{Outer temperature field in the continuous phase} By
using an outer expansion for the scaled temperature field in the
continuous phase
\begin{equation}
T =T_0 +T_1\epsilon + O(\epsilon^2),
\end{equation}
the energy equation for the outer temperature field in its leading
order can be obtained from Eq.(4) as follows
\begin{eqnarray}
1+U \frac{\partial{T_0}}{\partial{r}} +\frac{V}{r} \frac{\partial{T_0}}{\partial{\theta}}= 0.
\end{eqnarray}
By using the coordinate transformation from $(r,
\theta)$ to $(\psi_0,\theta)$ to solve Eq. (25), its solution can be written as
\begin{equation}
T_0(r,\theta) = G(\psi_0) - \int \frac{2r^4}{2r^3+1} \frac{d \theta}{\sin \theta},
\end{equation}
where $G(\psi_0)$ is a function of $\psi_0$ (the streamfunction in the
continuous phase).
Following\cite{6}, the solution near $r=1$ is simplified as
\begin{equation}
\begin{array}{ll}
T_0(r,\theta) =& (1+\frac{\pi}{6\sqrt{3}}-\frac{1}{6} \ln432) -\frac{1}{18} (\frac{\pi}{\sqrt{3}} +\ln432)
(r^2-\frac{1}{r}) \sin^2 \theta + \frac{1}{3} \ln(r^2-\frac{1}{r})\\
 & +\frac{2}{3}\ln(1+\cos \theta)
+ \frac{2}{9} (r^2 -\frac{1}{r}) \cos \theta +\frac{1}{9}  (r^2 -\frac{1}{r}) \ln(r^2 -\frac{1}{r}) \sin^2 \theta\\
& +\frac{2}{9}  (r^2 -\frac{1}{r}) \sin^2 \theta \ln(1+ \cos \theta).
\end{array}
\end{equation}
By using the boundary layer approximation
\begin{equation}
x= \frac{r-1}{\epsilon},
\end{equation}
the temperature field near interface can be expressed as
\begin{equation}
t(x,\theta)= 1+\frac{\pi}{6\sqrt{3}} -\frac{1}{6} \ln48
+\frac{2}{3} \ln (\frac{1 +\cos \theta}{\sin \theta}) + \frac{1}{3} x \sin^2 \theta \epsilon
\ln \epsilon + O(\epsilon).
\end{equation}

\subsection{Outer temperature field within the droplet}

By using the following outer expansion for the scaled temperature
field within the droplet in Eq. (23)
\begin{equation}
T' =\frac{1}{\epsilon^2} T'_{-2} + T'_0 + O(\epsilon),
\end{equation}
the equation in its leading order can be written as
\begin{equation}
\begin{array}{l}
U' \frac{\partial{T'_{-2}}}{\partial{r}} +\frac{V'}{r} \frac{\partial{T'_{-2}}}{\partial{\theta}}= 0.
\end{array}
\end{equation}
Its solution is
\begin{equation}
\begin{array}{l}
T'_{-2} = F_0(\psi'),
\end{array}
\end{equation}
where $\psi'=\frac{3}{4} \sin^2 \theta (r^4 - r^2)$ is the
streamfunction within the droplet. The unknown function $F_0(\psi')$
can be obtained from the following equation for the temperature
field $T'_0$ in its second order
\begin{equation}
U' \frac{\partial{T'_0}}{\partial{r}} +\frac{V'}{r} \frac{\partial{T'_0}}{\partial{\theta}}=\lambda \Delta F_0 + \frac{\lambda}{2\beta}.
\end{equation}

To solve Eq. (33), we use the coordinate transformation from $(r,
\theta)$ to $(m,q)$, with streamlines and their orthogonal lines
as coordinate axes defined as
\begin{equation}
\begin{array}{l}
m = -\frac{16}{3} \psi',\\
q = \frac{r^4 \cos^4 \theta}{2r^2-1}.
\end{array}
\end{equation}
Eq.(33) can thus be written in the $(m,q)$ coordinate system as follows
\begin{equation}
\frac{V_q}{h_q} \frac{\partial{T'_0}}{\partial{q}}
=\frac{\lambda}{h_m h_q h_{\phi}} \frac{\partial}{\partial m} (\frac{h_q h_{\phi}}{h_m} \frac{dF_0}{dm}) + \frac{\lambda}{2\beta},
\end{equation}
where $h_m$, $h_q$, $h_{\phi}$ are the metrical coefficients for
the transformation. Multiplying both sides of Eq.(35) by $h_m h_q
h_{\phi}$ and integrating with respect to $q$ for one circuit
along a streamline, we obtain the following equation
\begin{equation}
\oint V_q h_m h_{\phi}  \frac{\partial{T'_0}}{\partial{q}} dq
=\lambda \frac{\partial}{\partial m} [J(m) \frac{dF_0}{dm}] +\frac{\lambda}{2\beta} H(m),
\end{equation}
where $H(m)=\oint h_mh_qh_{\phi} dq$ and $J(m) =\oint \frac{h_q h_{\phi}}{h_m} dq$. From the equation
of continuity, it can be shown that $V_q h_m h_{\phi}$ is a
constant. We thus obtain a solution of Eq. (36) as follows
\begin{equation}
F_0= -\frac{1}{2\beta} [B + \int_0^m \frac{dx}{J(x)}\int_1^xH(s)ds].
\end{equation}
Near $m=0$, $J(m)$ and $H(m)$ can be expanded, respectively
\begin{equation}
J(m)= \frac{16}{3} -5m +O(m^2\ln m)
\end{equation}
and
\begin{equation}
H(m)= \frac{3}{8}\ln2 - \frac{1}{16} \ln m +O(m\ln m).
\end{equation}
Using Eqs. (38) and (39), we thus obtain from Eq. (37)
\begin{equation}
T'_{-2}(r, \theta)= F_0= -\frac{1}{2 \beta} [B - \frac{1}{16}m + \frac{3}{256}(3\ln2-1\frac{3}{4})m^2 -\frac{3}{512}m^2\ln m] + O(m^3\ln m).
\end{equation}
By using the boundary layer approximation
\begin{equation}
x'= \frac{1-r}{\sqrt{\lambda}\epsilon},
\end{equation}
the temperature field near interface can be expressed as follows
\begin{equation}
\begin{array}{ll}
t'(x', \theta) & =
\frac{\sqrt{\lambda}}{4 \beta}  x' \sin^2 \theta  \frac{1}{\epsilon}
+ \frac{3 \lambda}{16 \beta} x'^2 \sin^4  \theta  \ln \epsilon
+ o(\ln \epsilon).
\end{array}
\end{equation}

\subsection{ Inner temperature fields in the leading order}

From Eqs. (19) and (20), the leading orders of the inner temperature fields
in the continuous phase and within the droplet can be obtained, respectively.
A schematic diagram of the inner and outer regions with the leading orders 
near the surface of droplet is shown in Fig. 2.
By using inner expansions for the continuous phase and the fluid
in the droplet
\begin{eqnarray}
t(x, \theta)= t_{-1} \frac{1}{\epsilon}  + t_{l0} \ln \epsilon + t_0 + t_{l1} \epsilon \ln \epsilon   + O(\epsilon ),\\
t'(x', \theta)=  t'_{-1} \frac{1}{\epsilon} + t'_{l0} \ln \epsilon + t'_0 + t'_{l1} \epsilon \ln \epsilon   + O(\epsilon )
\end{eqnarray}
and the inner variables given in Eqs. (28) and (41), the scaled
energy equations for the  inner temperature fields in the leading
orders can be written as follows
\begin{eqnarray}
-3x \cos \theta \frac{\partial t_{-1}}{\partial x} + \frac{3}{2} \sin \theta \frac{\partial t_{-1}}{\partial \theta}
= \frac{\partial^2 t_{-1}}{\partial x^2},\\
-3x' \cos \theta \frac{\partial t'_{-1}}{\partial x'} + \frac{3}{2} \sin \theta \frac{\partial t'_{-1}}{\partial \theta}
= \frac{\partial^2 t'_{-1}}{\partial x'^2}.
\end{eqnarray}
The boundary conditions are
\begin{equation}
\begin{array}{l}
t_{-1}(0,\theta) =t'_{-1}(0,\theta),\\
\delta \frac{\partial{t_{-1}}}{\partial x}(0,\theta) = -\frac{\partial{t'_{-1}}}{\partial x'}(0,\theta),\\
t_{-1}(x \to \infty, \theta) \to 0,\\
t'_{-1}(x' \to \infty, \theta) \to B + \frac{\delta}{4} x' \sin^2 \theta,
\end{array}
\end{equation}
where $\delta=\sqrt{\lambda}/\beta$. We transform the
independent variables from $[(x,x'),\theta]$ to
$[(\eta,\eta'),\xi]$ and the functions from $(t_{-1},t'_{-1})$ to
$(f_0,f'_0)$ as
\begin{equation}
\begin{array}{l}
(\eta,\eta')= (\frac{3}{2} x \sin^2 \theta, \frac{3}{2} x' \sin^2 \theta),\\
\xi=\frac{1}{2}(2-3\cos \theta +\cos^3 \theta) = \frac{1}{2} (2+ \cos \theta)(1- \cos \theta)^2
\end{array}
\end{equation}
and
\begin{equation}
\begin{array}{l}
f_0(\eta,\xi)= t_{-1}(x, \theta),\\
f'_0(\eta',\xi)= t'_{-1}(x', \theta) - B  - \frac{\delta}{4} x' \sin^2 \theta.
\end{array}
\end{equation}
The corresponding energy equations for $f_0,f'_0$ and boundary
conditions can be written as follows
\begin{equation}
\begin{array}{l}
\frac{\partial f_0}{\partial \xi} =\frac{\partial^2 f_0}{\partial \eta^2},\\
\frac{\partial f'_0}{\partial \xi} =\frac{\partial^2 f'_0}{\partial \eta'^2}\\
\end{array}
\end{equation}
and
\begin{equation}
\begin{array}{l}
f_0(0, \xi)= f'_0(0,\xi) +B,\\
\delta \frac{\partial{f_0}}{\partial \eta}(0,\xi) = -\frac{\partial{f'_0}}{\partial \eta'}(0,\xi) -\frac{\delta}{6},\\
f_0(\eta \to \infty, \xi)= 0,\\
f'_0(\eta' \to \infty, \xi) =0.
\end{array}
\end{equation}
To solve Eq. (50), initial conditions are provided below
\begin{equation}
\begin{array}{l}
f_0(\eta,0) =0,\\
f'_0(\eta',0) =f'_0(\eta',\xi(\pi)) =f'_0(\eta',2) =g_0(\eta'),\\
g_0(\eta' \rightarrow \infty) \rightarrow 0.
\end{array}
\end{equation}
Following the methods given by Harper and Moore\cite{15}, the solution
of Eq.(50) can be obtained as
\begin{equation}
\begin{array}{ll}
f_0(\eta,\xi) =& \frac{1}{1+\delta} \{ (B-\frac{\delta}{6}\eta){\rm erfc}(\frac{\eta}{2\sqrt{\xi}})
+ \frac{\delta \sqrt{\xi}}{3 \sqrt{\pi}} \exp(- \frac{\eta^2}{4\xi}) \\
& + \frac{1}{\sqrt{\pi\xi}} \int_0^{\infty} g_0(\eta^*) \exp[-\frac{(\eta +\eta^*)^2}{4\xi}]d \eta^* \},\\
\end{array}
\end{equation}

\begin{equation}
\begin{array}{ll}
f'_0(\eta',\xi) =& \frac{\delta}{1+\delta} \{
 -(B +\frac{1}{6}\eta'){\rm erfc}(\frac{\eta'}{2\sqrt{\xi}}) + \frac{\sqrt{\xi}}{3 \sqrt{\pi}} \exp(- \frac{\eta'^2}{4\xi})\} \\
&+ \frac{1}{2\sqrt{\pi \xi}} \int_0^{\infty} g_0(\eta^*) \{ \exp[-\frac{(\eta'-\eta^*)^2}{4\xi}]
+ \frac{1-\delta}{1+\delta} \exp[-\frac{(\eta'+\eta^*)^2}{4\xi} ] \}d\eta^* .\\
\end{array}
\end{equation}
From Eqs.(49) and (53), we obtain the  inner
temperature field in its leading order near the surface of droplet
\begin{equation}
\begin{array}{ll}
t_{-1}(0, \theta)&= f_0(0,\xi) \\
&= \frac{1}{1+\delta} [B+  \frac{\delta \sqrt{\xi}}{3 \sqrt{\pi}} +\frac{1}{\sqrt{\pi \xi}} \int_0^{\infty} g_0(\eta^*) \exp(-\frac{{\eta^*}^2}{4\xi})d\eta^* ]\\
&= \frac{1}{1+\delta} [B+  \frac{\delta \sqrt{\xi}}{3 \sqrt{\pi}} +\frac{2}{\sqrt{\pi}} \int_0^{\infty} g_0(2 \xi^{1/2} \zeta) \exp(-\zeta^2)d\zeta].
\end{array}
\end{equation}
When the inner expansion in the temperature field (43) is truncated
at the zero order,  we rewrite Eq. (17) as
\begin{equation}
V_\infty = \frac{1}{2+3\alpha} \int_0^{\pi} \sin \theta \cos \theta [t_{-1} (0,\theta) \frac{1}{\epsilon}  +t_{l0} (0,\theta) \ln \epsilon] d\theta.
\end{equation}
Since $\epsilon=1/\sqrt{MaV_{\infty}}$, the migration speed of the droplet is evaluated as
\begin{equation}
V_\infty \approx a_1^2 Ma -2 a_{l0} \ln Ma + a_0,
\end{equation}
where
\begin{equation}
a_1 = \frac{1}{2+3\alpha} \int_0^{\pi} \sin \theta \cos \theta t_{-1} (0,\theta) d\theta
\end{equation}
and
\begin{equation}
a_{l0} = \frac{1}{2+3\alpha} \int_0^{\pi} \sin \theta \cos \theta t_{l0} (0,\theta) d\theta.
\end{equation}
Substituting Eq. (55) into Eq. (58), we obtain
\begin{equation}
\begin{array}{ll}
a_1 & =\frac{\delta}{3\sqrt{\pi}(2+3\alpha)(1+\delta)} \int_0^\pi \sin \theta
\cos \theta {\xi}^{1/2} d \theta\\
& +\frac{2}{\sqrt{\pi}(2+3\alpha)(1+\delta)} \int_0^{\pi} \sin \theta \cos \theta [\int_0^{\infty}
g_0(2 \xi^{1/2} \zeta) \exp(-\zeta^2)d\zeta] d\theta.
\end{array}
\end{equation}
To determine the function $g_0$ in Eq. (60), we use the boundary
condition within the droplet at the front and rear stagnation points
in Eq. (52)
\begin{equation}
\begin{array}{ll}
g_0(\eta') =& \frac{\delta}{1+\delta} \{-(B + \frac{1}{6} \eta'){\rm erfc}(\frac{\eta'}{2\sqrt{2}})
+ \frac{\sqrt{2}}{3\sqrt{\pi}} \exp(-\frac{\eta'^2}{8}) \}\\
&+ \frac{1}{2\sqrt{2\pi}} \int_0^{\infty} g_0(\eta^*) \{\exp[-\frac{(\eta'-\eta^*)^2}{8}]
+ \frac{1-\delta}{1+\delta} \exp[-\frac{(\eta'+\eta^*)^2}{8}] \}d\eta^*.
\end{array}
\end{equation}
The integral of the third
term on the right-hand side of Eq. (61) is approximated as
\begin{equation}
\int_0^\infty g_0(\eta^*) h(\eta',\eta^*) d\eta^* = \int_0^{\eta^*_l}g_0(\eta^*) h(\eta',\eta^*) d\eta^*
+ g_0(\eta^*_l) \int_{\eta^*_l}^{\infty} h(\eta',\eta^*) d\eta^*.
\end{equation}
 Then, Eq. (61) is evaluated
in a linear system of equations
\begin{equation}
\begin{array}{ll}
g_0(\eta') & - \frac{1}{4\sqrt{2\pi}} g_0(\eta^*_1)  \{\exp[-\frac{(\eta'-\eta^*_1)^2}{8}]
+ \frac{1-\delta}{1+\delta} \exp[-\frac{(\eta'+\eta^*_1)^2}{8}] \} \Delta \eta^*\\
& - \frac{1}{4\sqrt{2\pi}} g_0(\eta^*_{N+1})  \{\exp[-\frac{(\eta'-\eta^*_{N+1})^2}{8}]
+ \frac{1-\delta}{1+\delta} \exp[-\frac{(\eta'+\eta^*_{N+1})^2}{8}] \} \Delta \eta^*\\
& - \frac{1}{2\sqrt{2\pi}} \sum_{j=2}^N g_0(\eta^*_j)  \{\exp[-\frac{(\eta'-\eta^*_j)^2}{8}]
+ \frac{1-\delta}{1+\delta} \exp[-\frac{(\eta'+\eta^*_j)^2}{8}] \} \Delta \eta^*\\
& - \frac{1}{2} g_0(\eta^*_{N+1}) [ {\rm erfc} (\frac{\eta^*_{N+1}+\eta'}{2\sqrt{2}})
+  \frac{1-\delta}{1+\delta} {\rm erfc} (\frac{\eta^*_{N+1}+\eta'}{2\sqrt{2}}) ]\\
&= \frac{\delta}{1+\delta} [-(B + \frac{1}{6} \eta'){\rm erfc}(\frac{\eta'}{2\sqrt{2}})
+ \frac{\sqrt{2}}{3\sqrt{\pi}} \exp(-\frac{\eta'^2}{8}) ],\\
\end{array}
\end{equation}
where $\eta^*_{N+1}=\eta^*_l$ and $\Delta \eta^*= \eta^*_l/N$.
The physical coefficients used in the space experiment\cite{13} with the uniform temperature gradient
$G=12$K/cm for the continuous phase of Fluorinert
FC-75 and the droplet of 5cst silicon oil at $T=333$K are adopted here to yield
$\alpha=0.342$, $\beta=0.571$ and $\lambda=0.299$.
A typical value for $\eta^*_l$ is chosen as 3.
Using the trial
and error method to satisfy the above approximation, we determine
the unknown constant $B=0.1935$ and
obtain the dependence of $g_0$ on $\eta'$ as shown in Fig. 3.
From Eq. (60), we can determine the root-mean-square of the leading order term of the migration
speed as
\begin{equation}
a_1 =1.15 \times 10^{-2}.
\end{equation}
Although equations and boundary conditions describing the second order term of the migration speed
can be obtained, we are unable to find an analytical result for $t_{l0}$ in Eq. (56).
Under the truncation after the leading order term in Eq.(57), we obtain the migration speed of the droplet
\begin{equation}
V_\infty \approx 1.3 \times 10^{-4} Ma.
\end{equation}
By using the migration speed $V_\infty$, the adding thermal source $\Omega$ in Eq. (22)
can be determined as
\begin{equation}
\Omega = Gv_0 V_\infty (1 + \frac{\lambda}{2 \beta}) \approx 0.1 Ma K/s.
\end{equation}

\section{Conclusions and discussion}
\label{sec:sum} In this paper, the conservative integral
thermal flux across the surface for a steady thermocapillary droplet
migration in a uniform temperature gradient at large Re
and Ma numbers has been guaranteed by adding a
thermal source in the droplet. Under the assumptions
 of quasi-steady state and non-deformation of the droplet, we have made
 an analytical result for the steady thermocapillary
migration of droplet with the thermal source at large Re and Ma
numbers. The result shows that the thermocapillary droplet migration
speed slowly increases with the increase of Ma number.

We emphasize that the modeling of the adding thermal source in the droplet is
used to investigate the effect of integral thermal flux across the
surface on the thermocapillary droplet migration process.
It is one of the physical means to satisfy the conservative
integral thermal flux boundary condition in the
steady migration process at large  Ma numbers\cite{WH3}.
On the one hand, in general, when the droplet moves forward
under the uniform temperature gradient,
 the thermal energy in the droplet is imported from
the upper surface of the droplet  and exported from the lower surface.
The adding thermal source in the droplet can decrease/increase the
inner thermal flux from the interface/inside to the inside/interface at the upper/lower surface
of the droplet and satisfy the thermal flux boundary condition at the surface of
droplet for the steady migration process.
On the other hand,
at large Ma numbers, the heat advection around the droplet is
a more significant mechanism for heat transfer across/around the droplet.
 It is always keeping the weak transport of thermal
energy from outside of the droplet to inside.
Thus, the adding thermal source is conducive to providing the thermal energy in the droplet
to meet the requirement put forward by the quasi-steady migration process.
Meanwhile, the heat transport across the droplet surface becomes
weaker than the heat convection as Ma number increases. To reach the steady migration process,
more thermal energy in the droplet will be provided. 
The proportional relationship of the thermal source to Ma number in Eq. (66) qualitatively confirms the above
requirement of more thermal energy in the droplet.

In final, some suggestions are provided in order that the thermal source in the droplet
can be implementend in a real experiment. As given in Eqs. (1),(66) and shown in Fig. 1,
the thermal source depending on Ma number will
be added in the droplet when it starts to move under the uniform temperature gradient.
The heat radiant technology may be one of the possible physical means to add the thermal source
in the droplet.

\newpage
\textbf{Acknowledgments} We thank the National
Science Foundation for partial support through the Grant No.
11172310 and the IMECH/SCCAS SHENTENG 1800/7000 research computing
facilities for assisting in the computation.

\newpage
\textbf{Figure caption}

 Fig.~1. A schematic diagram of thermocapillary migration of droplet with a thermal source
 in a laboratory coordinate system.

 Fig.~2. A schematic diagram of the inner and outer temperature fields denoted by different colors
 near the surface of droplet in a coordinate system moving with
the droplet velocity $V_{\infty}$.
  The light/dark gray and the white/black regions represent
 the inner [$t/t' \sim 0(\frac{1}{\epsilon})/0(\frac{1}{\epsilon})$] and
 outer [$T/T' \sim 0(1)/0(\frac{1}{\epsilon^2})$]) temperature fields
 in the continuous phase/within the droplet, respectively.

 Fig.~3. Function $g_0$ versus $\eta'$ determined from Eq. (63).

\end{document}